\documentclass[reprint,amsmath,amssymb,aps,prl,showpacs,superscriptaddress]{revtex4-1}

\usepackage[utf8]{inputenc}
\usepackage{graphicx}
\usepackage{dcolumn}
\usepackage{bm}
\usepackage{textcomp}
\usepackage[colorlinks,linkcolor=blue,citecolor=blue]{hyperref} 

\def\be{\begin{equation}}

\def\ee{\end{equation}}
\def\barr{\begin{array}}
\def\earr{\end{array}}

\def\ed{\end{document}}

\begin{document}

\title{Correlation networks from random walk time series}

\author{Harinder Pal}
\affiliation{Instituto de Ciencias F\'isicas, Universidad Nacional Aut\'onoma de M\'exico, Cuernavaca, Morelos - C.P. 62210, M\'exico} 
\author{Thomas H. Seligman}
\affiliation{Instituto de Ciencias F\'isicas, Universidad Nacional Aut\'onoma de M\'exico, Cuernavaca, Morelos - C.P. 62210, M\'exico} 
\affiliation{Centro Internacional de Ciencias A. C., Cuernavaca, Morelos, M\'exico} 
\author{Juan V. Escobar}
\affiliation{Instituto de F\'isica, Universidad Nacional Aut\'onoma de M\'exico, PO Box 20-364, M\'exico City, 04510, M\'exico}

\date{\today}

\begin{abstract}
Stimulated by the growing interest in the applications of complex networks framework on time series analysis, we devise a network model in which each of $N$ nodes is associated with a random walk of length $L$. Connectivity between any two nodes is established when the Pearson correlation coefficient(PCC) of the corresponding time series is greater than or equal to a threshold $H$, resulting in similarity networks with interesting properties. In particular, these networks can have high average clustering coefficients, ``small world" property, and their degree distribution can vary from scale-free to quasi-constant depending on $H$. A giant component of size $N$ exists until a critical threshold $H_c$ is crossed, at which point relatively rare walks begin to detach from it, and remain isolated. This model can be used as a first step for building a null hypothesis for networks constructed from time series.           
\end{abstract}
 


\maketitle

\section{I. Introduction}
\label{Intro}
Networks are mathematical abstractions that help us understand 
complex interactions between elements. They constitute a mapping in which such elements are represented by nodes and their interactions are represented by ``edges" or ``links". In the last couple of decades, the study of complex networks \cite{Albert, Barrat} has gained much interest, as a wealth of systems spanning many and diverse fields of research have benefited from it, including Epidemiology \cite{Pastor}, Neuroscience \cite{Stam},  Economics \cite{Schweitzer}, Linguistics \cite{Motter}, Biology \cite{Kauffman,Montoya} , Physics \cite{Dorogovtsev}, Information theory \cite{Franceschetti} and, of course, Social Sciences \cite{deSolaPool}, among others. 

The widespread availability of data in the internet age has been driving force for the boom in complex network research. In particular, complex network theory has recently found many useful applications in the study of time series analysis \cite{Zhong}. In some cases, mapping a single time series onto a network and a careful study of its resulting statistics can reveal intrinsic features of the time series, particularly whether the signal is periodic, pseudoperiodic or multifractal \cite{Lacasa, Budroni, Zhang}, and even be used for the diagnosis of diseases \cite{Small}. 

A different approach has been to study the relation between distinct time series by assigning a node to each one of them, along with some criterion to establish when a link is present. In this respect, whether nor not two nodes are liked in many real-life networks is unambiguous, as it usually follows directly from a definition. For example, in the actors' network, a link is formed between two actors if and only if they worked in the same movie \cite{Barabasi}; in the WWW network, two servers are considered to be linked if one of them has a webpage that contains a link to a webpage on the other \cite{Jeong}; in some semantic networks, two words are linked if they are synonyms according to a particular dictionary \cite{Ravasz}. However, when applied to time series, the links joining the network may not be so trivially established. One way to establish the connection between nodes is through the correlation of the corresponding time series. For example, Zhang and Small have studied pseudoperiodic time series by constructing complex networks by assigning portions of the time series to nodes which are linked on the basis of correlations between them \cite{Small}. Y. Yang and H. Yang \cite{Yang} have also used the correlation-based approach to construct complex networks from stock time series and suggest a strategy to choose an appropriate critical value (or threshold $H$) of the correlation for translating the correlation matrix into the adjacency matrix. Tang at al, \cite{Tang} also have used correlations in the study of traffic time series, as well as thresholds to define when a link is established. Using correlations to establish links in this way implies that the resulting network may be regarded as a sort of ``similarity" one, in which the threshold indirectly defines how much mutual information two time series must have in order to be connected \cite{Reza}. In this context, the natural question of precisely which threshold may be considered as appropriate to establish a link is inherently related to the fact that even time series that arise from some random process will in general have a correlation different than zero. A similar issue is present regarding the simpler real-world networks mentioned above: can the observed properties arise from pure randomness? The random network or random graph model, popularly known as Erd\H{o}s and Renyi's network \cite{Erdos}, is considered as the first step towards the understanding of real complex networks, and serves as a null hypothesis for their origin. Actually, out of the most frequent properties of real networks (scale-free degree distribution, large clustering coefficient, the existence of a giant component, the so called ``small world'' property, and the presence of hierarchical structures), random graphs can reproduce the existence of a giant component (in the super-critical regime) and the small world property, which is assigned when the average shortest path between nodes is small and grows at the most as the $Log(N)$.

In this work we study the properties of the equivalent random system, but for networks arising from time series: the undirected networks in which each of $N$ nodes is associated to a uniform step size random walk of length $L$, and links between nodes are assigned when the Pearson correlation coefficient (PCC) between the corresponding walks is above a threshold $H$. Analogously, we propose that our model can be regarded as a null-hypothesis for networks created from time series. 

Despite the simplicity of the model, we find interesting properties as a function of $H$. In particular, the resulting networks are extremely robust: below a threshold $H_c$, there is giant component of size $N$, but, unlike random graphs, the expelled nodes remain isolated for higher thresholds. Furthermore, the networks are small world and have very large clustering coefficients, except for $H$ close to 1. Also, the degree distribution can be tuned from quasi-constant to power law. We also analyze the degree correlations and clustering coefficient spectra as a function of $H$. Our model resembles in spirit the ``hidden variable" model proposed by Caldarelli et. al \cite{Caldarelli} later generalized by Bogu\~{n}\'{a} and Pastor-Satorras \cite{Boguna}, but differs from it in that there is no \textit{a priori} single variable (or fitness) associated with each node that could be drawn from some distribution function. However, we briefly show that there exists a mapping from the time series onto a single random variable (the final position of the walk) that, along with a relatively simple threshold rule, yields qualitatively similar results and is amenable to potential analytic treatment following the general treatment developed in \cite{Boguna}. 

Finally, it must be noted that a random walk is a non-stationary process. Many times, especially in stock analysis, a meaningful correlation coefficient between two time series can only be obtained after mathematically transforming them into stationary time ones \cite{book}. In the case of the random walks that we consider here, the corresponding transformation would entail replacing the original time series of distances with a time series of the first differences of the distances, a procedure referred to as detrending. However, in this work we limit ourselves to analyzing the properties of the networks arising from the random walk time series since, in many other instances, the trends in the time series are actually relevant and a calculation of the PCC is useful.
\section{II. The Model}
\label{II. model}
We consider $N$ one-dimensional random walks of length $L$ such that each random walker starts from the origin and takes a step of size one, either in the positive or in the negative direction. We calculate the Pearson's correlation coefficient between two random walk trajectories (RWT) with indices $i$ and $j$, denoted by $C_{ij}$, from the corresponding time series of positions $r^i_t$ and $r^j_t$, defined as 
\begin{equation}
C_{ij} = \dfrac{< r^i_t r^j_t > - <r^i_t><r^j_t>}{\sigma_i \sigma_j},
\label{eqPCC}
\end{equation}
where $<..>$ represents the time average and $\sigma_i$ and $\sigma_j$ are standard deviations of each walk. An undirected network is created in which a link between two nodes (representing two random walks) exists if the absolute value of the PCC between them is greater than or equal to a threshold $H$. To implement this, the adjacency matrix elements are defined as:
\begin{eqnarray}
A_{ij} &=& 1 \quad \mbox{if} \quad |C_{ij}| \geq H, \forall \quad i \ne j, \nonumber \\
       &=& 0 \quad \mbox{if} \quad |C_{ij}| < H, \forall \quad i \ne j, \nonumber \\
       &=& 0 \quad \forall \quad i = j
\end{eqnarray}
\begin{figure}
\vspace{0.2cm}\includegraphics[width=3.0in]{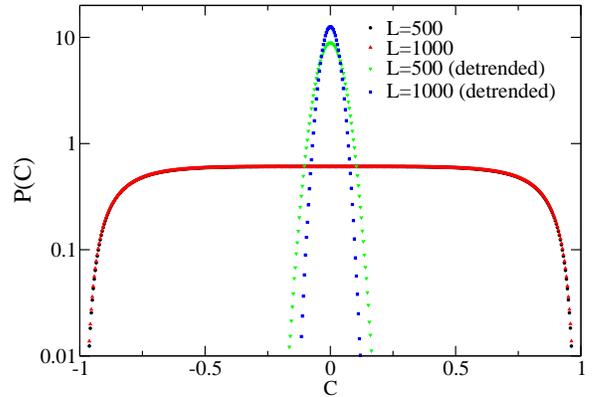} 
\caption{(color online) Normalized distributions of frequency $f$ vs $C$ for different values $L$ for original and detrended time series of random walks.}
\label{fig-corr-dist}
\end{figure}

Figure \ref{fig-corr-dist}, shows the normalized PCC distributions for different values of $L$ sufficiently large as to yield practically continuous probability distributions of $C_{ij}$. When this condition is fulfilled, the distribution does not depend on $L$, is practically constant for $|C|<0.5$, and decays exponentially after $|C|>0.875$. As a result, only about $2.5\%$ of the total walks present an absolute correlation larger than this latter value, and $0.3\%$ for $|C|>0.93$. Due to computational considerations, we fix $N=5000$ and average 100 different samples to reduce statistical noise. For the sake of completeness, the distribution of the correlation for the detrended data is also shown in Fig. \ref{fig-corr-dist}, which, in contrast to the original walks, is normally distributed around $0$. The resulting network for the detrended time series is simply a random network with a connection probability that depends on $L$. From hereon, we will refer only to the results of the original random walk series.
\section{III. Results and Discussion}
\label{Results and Discussion}
\section{IIIa. Average degree and degree distribution}
\label{IIIa.  Average degree and degree distribution}
For $H=0$, clearly the network is fully connected, $\quad A_{ij} =1 \quad \forall \quad i\ne j$, while, for $H=1$, the probability ($|C_{ij}|=1$) $\sim 0$ as it would require two RWTs to be either identical or reflections of each other. So $A_{ij}=0$ for $H=1$ for all but a vanishing small number of nodes for any given sample, which means almost all nodes are isolated. For intermediate values of $H$, the connections between different nodes slowly break on increasing $H$, as the fraction of correlation matrix elements for which $|C_{ij}| > H$ decreases and the resulting adjacency matrix becomes more and more sparse.

\begin{figure}[htb] 
\includegraphics[width=3.0in]{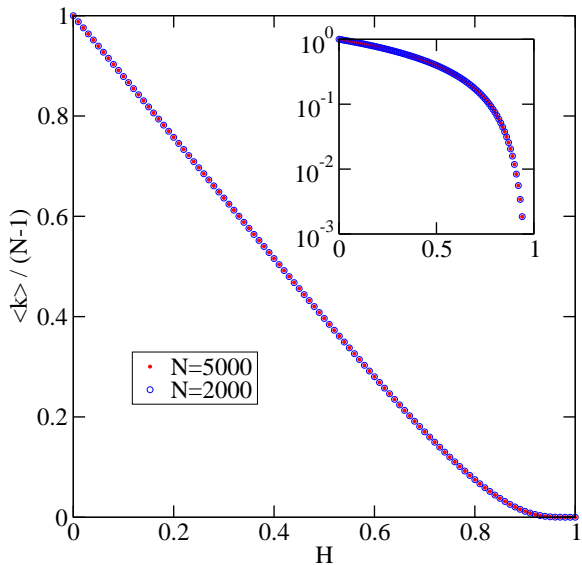} 
\caption{Normalized average degree vs $H$ for $N=5000$ and $L=500$ averaged over 100 samples.}
\label{fig-degree-vs-H}
\end{figure}

Since the network is fully connected for $H=0$, all the nodes have ($N-1$) links. As shown in Fig. \ref{fig-degree-vs-H}, from then on the average degree ($<k>$) decreases approximately linearly with $H$ up to $H=0.6$, and transitions to an exponential decay for $H>0.875$, which corresponds to the point at which the distribution of correlations also decreases exponentially in Fig.  \ref{fig-corr-dist}.
\begin{figure}
\hspace*{-0.25cm}\includegraphics[width=3.2in]{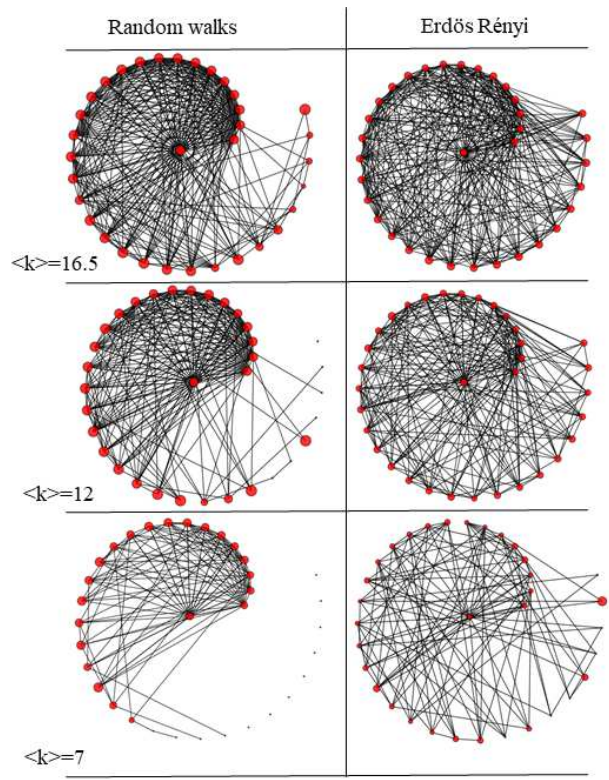} 
	\caption{Representation of the network and its evolution for $N=35$. (Left panel) Networks generated with the random walks and $L=500$ in which nodes are ordered in descending order of degree from the center of the spiral, and node size is proportional to its clustering coefficient for $H=0.4$ (top, $<k>=16.5$), $H=0.5$ (middle, $<k>=12$) and $H=0.65$ (bottom, $<k>=7$). The middle graph is close to the $H_c$, so that a few nodes are already disconnected from the giant component. (right panel) Networks generated using the Erdos Renyi model with the same $<k>$ as the networks from the left panel.} 
\label{fig-random-walk-spiral}
\end{figure}

The topology of the networks formed as well as its difference with that of random networks can be better grasped by looking at a graphical representation of its behavior $vs.$ $H$.  In Fig. \ref{fig-random-walk-spiral} we show an example of this evolution for $N=35$ and $L=500$ (larger number of nodes makes visualization challenging) for three different values of $H$, in which nodes are arranged in descending number of links starting from the center of the spiral outwards, and the size of the nodes are proportional to their clustering coefficient. Note that even the least connected nodes are connected to adjacent nodes, but also to the more popular ones at the center. This accounts for both the large clustering coefficients as well as for the  short average path lengths we find for these networks (see next section). These characteristics that arise naturally in our model, are also present in the one put forward by Watts and Strogatz \cite{Watts}. For comparison, the equivalent representation for random networks with the same average degrees is shown in the right panel of Fig. \ref{fig-random-walk-spiral}.   An animation of this evolution but for a larger number of threshold values can be found at \cite{animation1} and can be contrasted with a second animation at \cite{animation2} of an equivalent random network as the connection probability is reduced all the way from 1 to 0.

\begin{figure}[htb] 
\vspace{1.7cm}\includegraphics[width=3.1in]{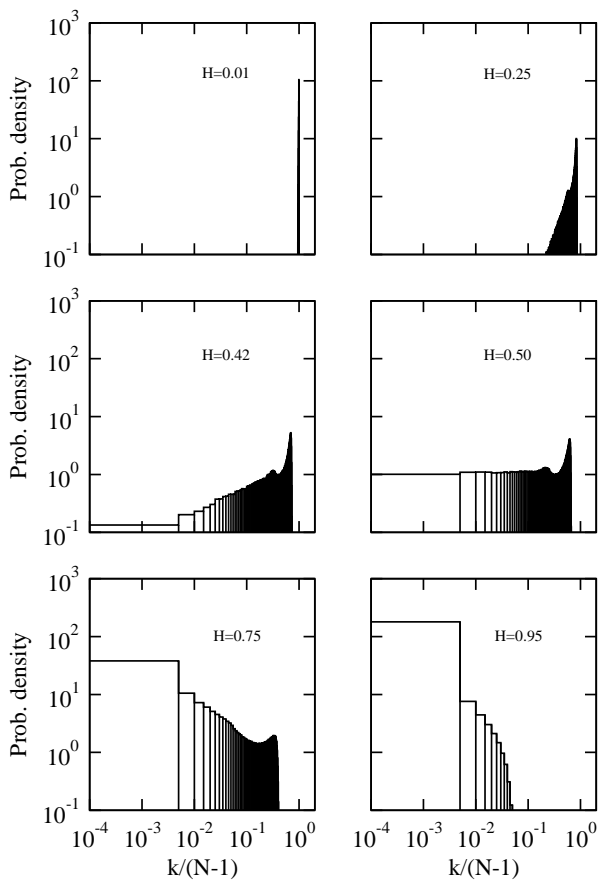} 
\caption{Degree distribution for different values of $H$ for $N=5000$ and $L=500$ averaged over 100 samples.}
\label{fig-degree-dist}
\end{figure}

Figure \ref{fig-degree-dist} shows the degree distribution for different values of $H$. Since in general different nodes have different distributions ($\rho(|C|)$) of correlations with other nodes, they also have a different tendency to maintain or lose connection as $H$ changes.  As a result, the degree distribution becomes more and more diffused until the network finally begins to disconnect at approximately $H=0.42$ (this exact point may vary with $N$ or the number of samples), where the distribution touches $k=0$. On further increasing $H$, the degree distribution in the bulk part becomes almost constant (for example for $H=0.5$). At larger values of $H$, the network becomes highly disconnected, and the probability for lower degrees becomes higher as shown, for example, for $H=0.75$. For this threshold, the distribution of degrees of the connected nodes for these large thresholds resembles a power law, except for a small hump present at the end of the distribution. A rationalization for the negative slope for large $H$ for the degree distribution is given in at the end of section IIIC, while an explanation for the presence of the two peaks observed for $H=0.42$ and $H=0.50$ is provided in the Appendix A.
\section{IIIb. Size of largest component, clustering coefficient and average shortest path}
\label{connectivity}
\begin{figure}
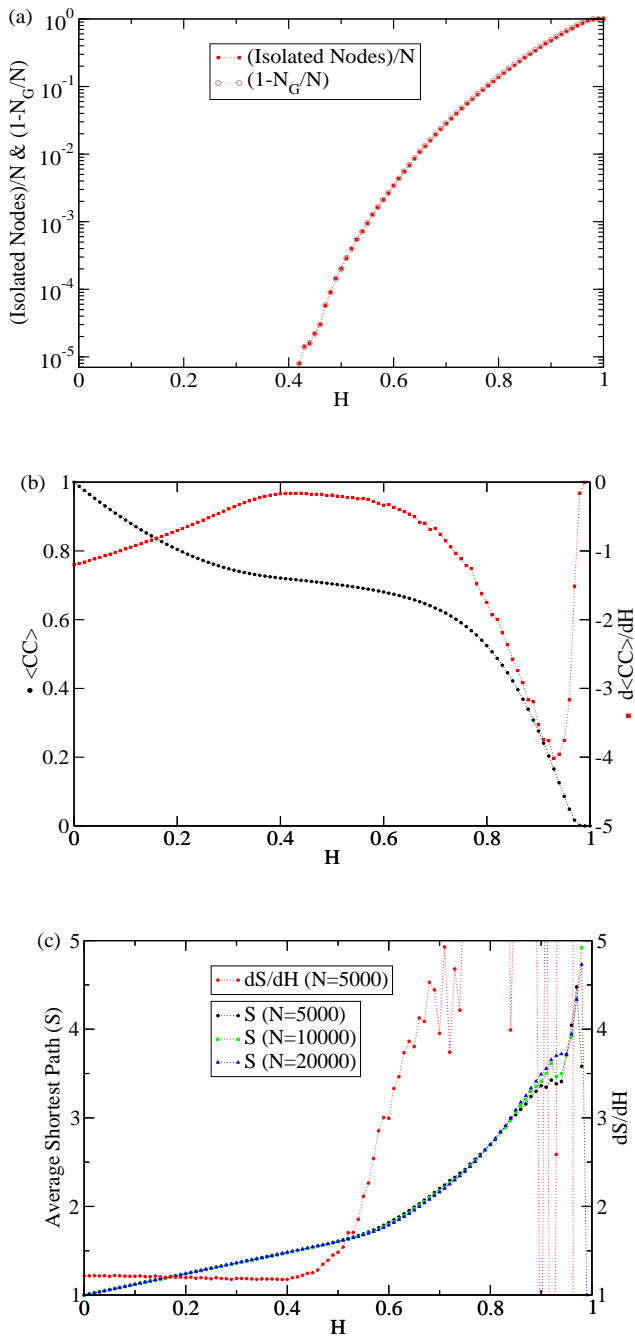

\hspace*{-0.65cm}\includegraphics[height=2.1in,width=3.15in]{figure5a.eps}\vspace{0.85cm} 
\hspace*{-0.15cm}\includegraphics[height=2.05in,width=3.22in]{figure5b.eps}\vspace{0.84cm}
\includegraphics[height=2.12in,width=3.1in]{figure5c.eps}\vspace{-0.2cm}
\caption{(a) $1-N_G/N$ and fraction of isolated nodes vs $H$ for $N=5000$ (averaged over 100 samples), (b) average clustering coefficient vs $H$ for $N=5000$ (averaged over 20 samples) and its derivative, (c) mean shortest path vs $H$ for different values of $N$ (one sample was used for each N) and its derivative for $N=5000$.}
\label{fig-SG-Clus-coef-and-mean-path}
\end{figure}

An interesting feature of the network created is that it is extremely robust. It remains a single component network up to a critical value of $H$, $H_c\approx0.42$. Thereafter, the nodes slowly begin to dissociate from the network as $H$ increases, but never into smaller sub-components of significant sizes, $i.e.$, detached nodes become and remain isolated. This is evidenced by plotting $(1- N_G/N$), where $N_G$ stands for the ``size of the giant component" simultaneously with the fraction of the isolated nodes ($F$) as shown in Fig. \ref{fig-SG-Clus-coef-and-mean-path}(a). A non-zero value of $(1-N_G/N)$ or $F$ smaller than $1/N$ would mean that the dissociation of nodes has taken place only in some of the averaged samples, which is responsible for noisy portion of $F$ and $(1-N_G/N)$ \textit{vs.} $H$ just after they become non-zero. Note also, that even for relatively high thresholds, the size of the largest component is quite large (\textit{e.g.}, $N_G/N \sim 0.85$ for $H=0.8$).

$H_c$ can be interpreted as the threshold at which redundancy in the connectivity of $N$ nodes is at a minimum, \textit{i.e.}, the threshold that yields the minimum number of links among the original $N$ nodes before the nodes begin to detach. The behavior found for $N_G$ contrasts with that exhibited by random networks, which not only present a phase transition at an average degree $\langle$k$\rangle=1$, but also never reach a giant component of size $N$. Here, there is no small number of hubs whose removal would separate the $N_G$ into smaller units of significant size. This is a consequence of the fact that these are effectively similarity networks, due to which the probability of a single walk linking two dissimilar groups of similar walks is practically zero (for large $N$). A second direct consequence of being similarity networks is the notably high average clustering coefficients displayed ($<CC>$, Fig. \ref{fig-SG-Clus-coef-and-mean-path}(b)). Indeed, if a walk $A$ is correlated with walks $B$ and $C$ for some $H$, then the probability that walks $B$ and $C$ are also correlated is high. A second characteristic of real networks also displayed by our model is that they are small world: Fig. \ref{fig-SG-Clus-coef-and-mean-path}(c) shows the average shortest path ($S$) only exceeds 4 for $H\approx1$, and, furthermore, that its shape \textit{vs} $H$ as well as its general magnitude remains invariant irrespective of $N$ for the different network sizes tested. Crossing the critical threshold also affects these two quantities: at $H_c$, the slope of $<CC>$ \textit{vs.} $H$ reaches a maximum (Fig. \ref{fig-SG-Clus-coef-and-mean-path}(b) (right axis)), while that of $S$ increases (Fig. \ref{fig-SG-Clus-coef-and-mean-path}(c) (right axis)). This suggests that $H_c$ separates two different network topologies as reflected in these two important quantities.
We point out that the exact critical threshold $H_c$ varies slightly with $N$ while remaining in the vicinity of $0.4$. In fact, we have numerically observed that H increases with N for a fixed $F$ or $(1 - N_G/N )$ away from both zero and one. This could be a result of the fact that the maximum value of $|C_{ij}|$ for a given node $i$ is likely to increase with $N$, even though the overall average distribution is independent of $N$. However, we cannot establish a definite trend in $H_c$ as a function of $N$. Nevertheless, the variations around $H_c=0.42$ are relatively small and become even smaller for larger $N$. 
\section{IIIc. Hidden variable \& fitness value mapping}
\label{Hidden-variable}
\begin{figure}
\hspace*{-0.8cm}\vspace*{0.6cm}\includegraphics[width=3.0in,angle=0]{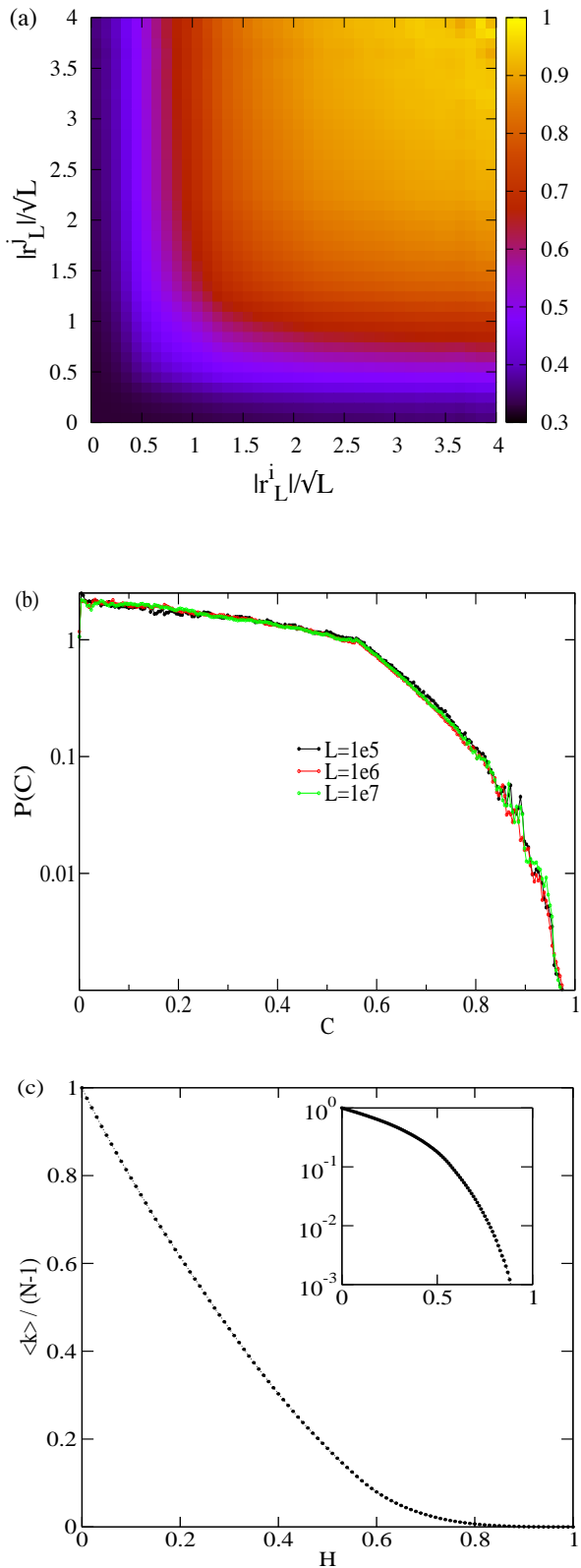}\vspace{0.9cm}
\hspace*{-0.35cm}
\includegraphics[trim=1cm 1cm 0cm 1cm, height=2.1in,width=2.9in]{figure6b.eps}\vspace{0.9cm}
\hspace*{-0.52cm}\includegraphics[height=2.6in,width=3.0in]{figure6c.eps}
\caption{(a) Average correlation on $({|r^i_L|}/{\sigma},{|r^j_L|}/{\sigma})$ plane. $N=5000$, $L=500$, number of samples=100. (b) Distribution of equivalent correlations for $L=10^{5},10^{6}$ and $10^{7}$. (c) Average degree vs $H$ using $N=5000$ and $L=10^{6}$. }
\label{fig-av-abs-PCC-Rl-plane}
\end{figure}

Given the general impact that the hidden variable model \cite{Caldarelli} has had in the field, as well of the general method to find analytical expressions for its relevant statistical quantities \cite{Boguna}, it is desirable to asses if the particular results we have found could be obtained --at least qualitatively -- within the framework of the former. This would entail substituting each of the Random walk time series (RWS)  with a single random variable, and finding a suitable function to play the role of the PCCs. To this end, we first note that two walks have a high probability of being correlated if the final distances $|r_L|$ are large compared to the standard deviation of the process, $\sqrt{L}$, since a relatively small number of combinations of forward and backward steps will bring them to those positions. Thus, if the final positions of two long distance-walks are similar, chances are they are highly correlated. Conversely, relatively short walks tend to be poorly correlated with any other walk even if their final positions are equal, as shown in Fig. \ref{fig-av-abs-PCC-Rl-plane}(a).  This suggests that $|r_L|/\sqrt{L}$ may play the role of an intrinsic fitness factor. With this in mind, we propose that each node $i$ be assigned a fitness value equal to $|r_L^i/\sqrt{L}|=|2n_i-L|$, where $n_i$ is the number of positive steps drawn from the probability distribution  $\rho (n,L)={{L}\choose{n}}  /2^{L}$. Then, we propose that the Pearson correlation function can be replaced by an equivalent function of $(r_L^i, r_L^j)$ of the form:
\begin{equation}
\begin{array}{rcl}
C^{eq}_{ij} & = & \Bigg\{1- \dfrac{|(|r_L^i|-|r_L^j|)|}{\alpha \sqrt{L}} \Bigg\} \\ 
& \times & \Bigg\{1-\Big( P_{rel}(r_L^i)P_{rel}(r_L^j)\Big)^{\beta}\Bigg\} \;,
\end{array}
\label{eqEquivCorrel}
\end{equation}   
where $P_{rel}(r_{Li})=e^{-((L/2)-n(r_L^i))^{2}/(L/2)}=e^{{-{r_L^i}^2}/{2L}}$ is the relative probability of obtaining a walk ending at $r_{Li}$ compared to a walk that ends at 0, and $\alpha$ and $\beta$ are constants. Eq. \ref{eqEquivCorrel} is a first order approximation to the observations made above. The first term on its \textit{rhs} is a measure of how close the normalized final positions of two walks are. The second term in this equation provides a non trivial-- and certainly non unique either-- correction to the first term aimed at minimizing its contribution for short-distance walks ($r_L^i/\sqrt{L} \approx0$, $P_{rel}(r_L^i/\sqrt{\sqrt{L}})\approx 1\implies C^{eq}_{i,j}\approx 0$), and maximizing it for long-distance ones ($r_L^i/\sqrt{L}\approx 1$, $P_{rel}(r_L^i/\sqrt{\sqrt{L}})\approx 0\implies C^{eq}_{ij}\approx 1$). 

\begin{figure}
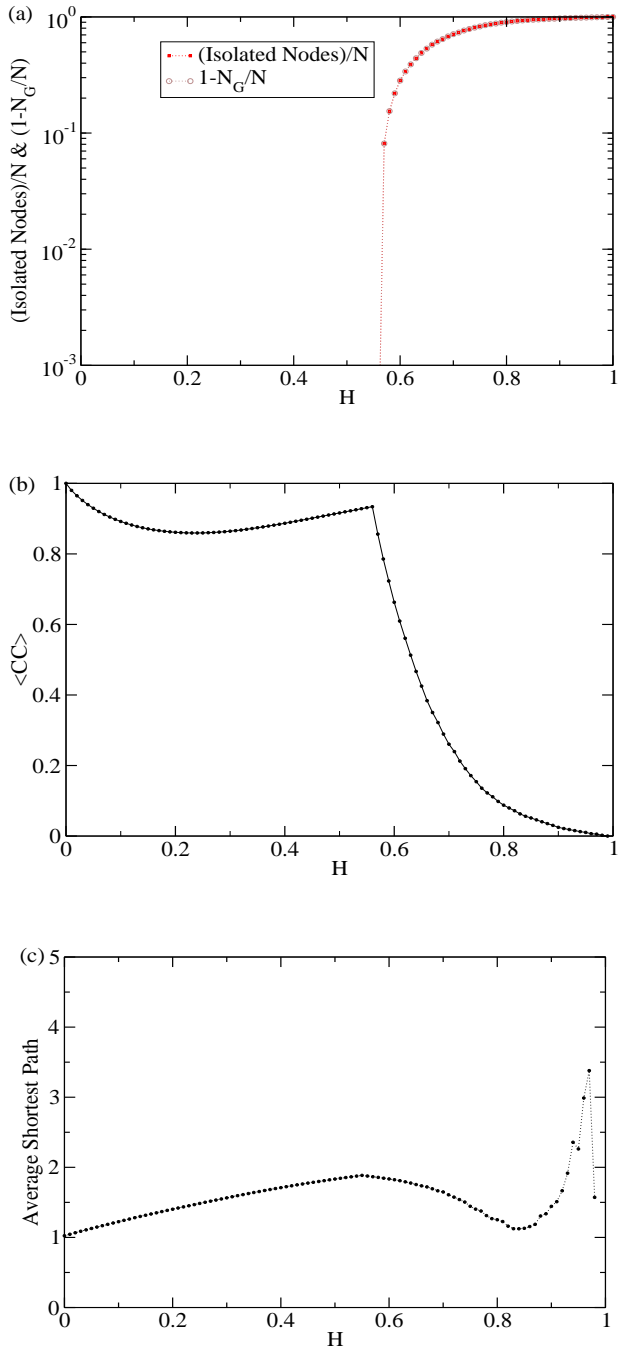

\hspace*{-0.65cm}\includegraphics[height=2.1in,width=3.2in]{figure7a.eps}\vspace{0.9cm}
\hspace*{-0.65cm}\includegraphics[height=2.1in,width=3.2in]{figure7b.eps}\vspace{0.9cm}
\hspace*{-0.6cm}\includegraphics[height=2.1in,width=3.1in]{figure7c.eps}
\caption{Statistical properties of an equivalent hidden-variable model. (a) $1-N_G/N$ and fraction of isolated nodes vs $H$, (b) Average Clustering Coefficient and (c) Average shortest path length for one sample with N=5000 and $L=10^{5}$. The qualitative features of the RWS's are recovered.}
\label{fig-Hidden-Variable-Cont} 
\end{figure}

By inspection, we find that $\alpha=8$ and $\beta=1/2$ yield a distribution of $C^{eq}_{ij}$ (see Fig. \ref{fig-av-abs-PCC-Rl-plane}(b)) for large L that resembles closely that obtained for the RWSs using the PCCs (Fig. \ref{fig-corr-dist}). It is worth pointing out that while the distribution of correlations is invariant \textit{wrt} $L$ as that of RWSs, unlike the latter, for this hidden variable model $L$ must be larger than $10^{4}$ for the distribution to be devoid of peaks. If now the linking probability function is given by
\begin{equation}
f(r_L^i,r_L^j)=\Theta(C^{eq}_{ij}-H) \;,
\label{eqConectionProb}
\end{equation} 
then Fig. (\ref{fig-Hidden-Variable-Cont}) shows that the main features of the networks obtained from random walks and PCC are recovered, albeit qualitatively. Specifically, there is a critical threshold (this time around $H=0.58$) below which $N_G/N=1$, while above it, the expelled nodes remain isolated (Fig. \ref{fig-Hidden-Variable-Cont}(a)). Furthermore, $<CC>$ is very large (Fig. \ref{fig-Hidden-Variable-Cont}(b)), and the average shortest path is small (Fig. \ref{fig-Hidden-Variable-Cont}(c)), along with the fact that its maximum value is about 4 for large $H$. Also, $<k>$ decreases with $H$ (Fig. \ref{fig-av-abs-PCC-Rl-plane}(c)) at a comparable rate as for the RWS system (Fig. \ref{fig-degree-vs-H}), upuntil the critical value. These results suggest that the relevant properties of the random walks that yield the corresponding networks are indeed captured by eqs. \ref{eqEquivCorrel} and \ref{eqConectionProb}. Then, it may be possible to derive all these and other properties analytically using the formulation developed in \cite{Boguna}. Having said this, we realize there are some clear differences between the results of RWSs and the equivalent hidden variable one. For example, $S$ decreases at $H_c$, and only increases again at higher values of $H$. Further improvements of this mapping may minimize these differences.

The analysis presented in this section also helps understand the negative slope of the bulk of the degree distribution for large $H$ (Fig. \ref{fig-degree-dist}): it can be attributed to the combined effect of the normalized fitness factor $r_L/\sigma$ and the exponential decay in its probability density as $P(r_L)=({2}/{\sqrt{2\pi L}}) e^{-{r_L}^2/{2 L}}$. Figure \ref{fig-av-abs-PCC-Rl-plane}(a) shows that the effect of $r_L$ is more profound in the regions (if one divides that plane into horizontal or vertical strips) in which the average correlations are higher. It effectively means that the statistics of the links that survive upon imposing a high threshold should be governed mainly by $r_L/\sqrt{L}$. Although $\rho(|C|)$ is different for different nodes, there do exist two groups of nodes such that the correlation distributions overlap significantly for nodes within these groups as explained in more detail in Appendix A. As a result, the connection probabilities, and hence the degrees for the nodes within each of these group are close to one another. Thus, there exist two peaks in the degree distribution for a wide range of $H$. These peaks can be treated as a manifestation of the random network-like behavior, to some extent, within these groups. 
\section{IIId. Degree correlation and clustering spectrum}
\label{connectivity}
\begin{figure}
\includegraphics[width=3.2in]{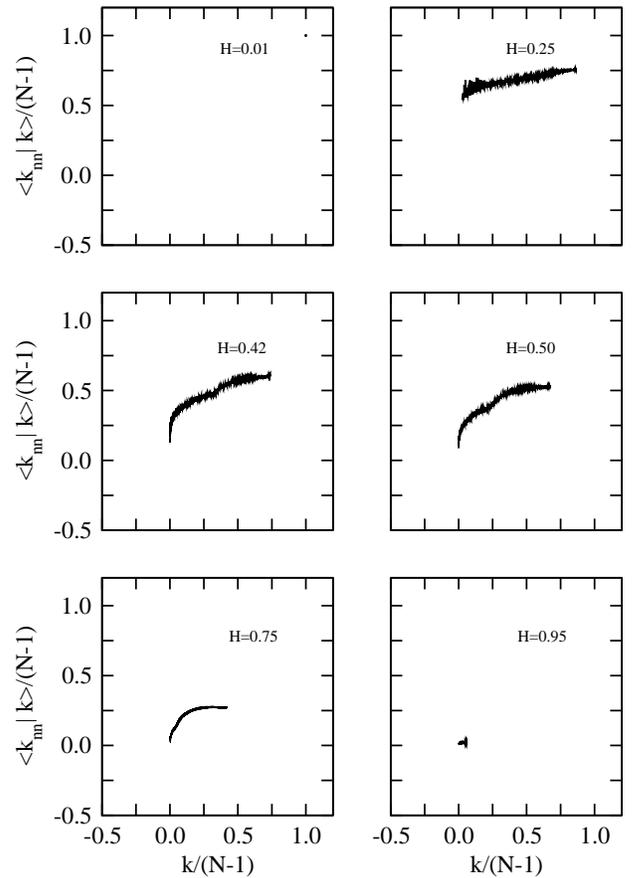} 
\caption{Degree correlation for different values of $H$ for $N=5000$ and $L=500$ averaged over 100 samples.}
\label{fig-degree-corr}
\end{figure}

\begin{figure}
\includegraphics[width=3.2in]{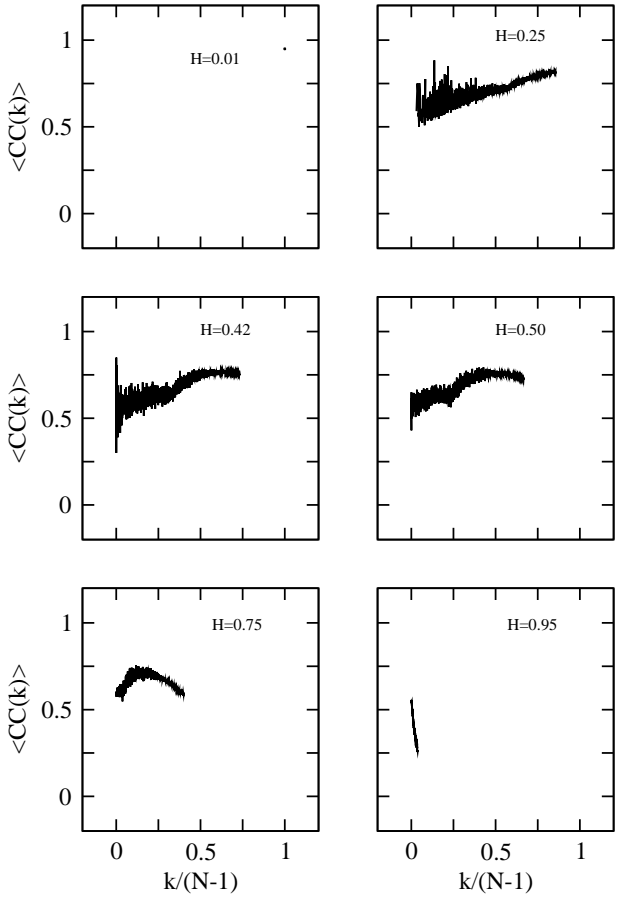} 
\caption{Clustering coefficient vs degree for different values of $H$ for $N=5000$ and $L=500$ averaged over 20 samples.}
\label{fig-clust-coeff-vs-k}
\end{figure}

The networks obtained from random walk time series exhibit an overall assortative degree correlation as shown in Fig. \ref{fig-degree-corr}. On the other hand, an analysis of the average clustering coefficients \textit{vs} $k$ shown in Fig. \ref{fig-clust-coeff-vs-k} reveals that these networks are conformed by definite structures, unlike networks formed with the Erd\H{o}s and Renyi or the Barabasi-Albert models for which $<C(k)>$ is constant. We find an overall positive correlation between $<C>$ and $k$ for a wide range of $H$ that becomes negative only for relatively large values of $H$.  The overall shape of the degree correlation and the presence of discontinuities in the slopes of  $<k_{nn}|k>$ and $<C(k)>$ vs $k$ are rationalized in Appendix B.  
\section{IV. Conclusions}
In summary, the model we have presented forms highly connected networks from random walk time series that do not disintegrate into components of comparable sizes as the threshold that determines linking is increased. Rather, the expelled nodes remain isolated. In the range $0.5 < H < 0.85$, the network possesses some important characteristics of real networks. Specifically, they have the small world property, high average clustering coefficient as well as a power law degree distribution with some deviations. Also, there exists a negative correlation between clustering coefficient and k, at least for high degrees. Furthermore, it is possible to reproduce, albeit qualitatively, the main threshold-dependent properties of these networks by substituting the walks assigned to the nodes with just their normalized final positions and an equivalent correlation function of these fitness values.

We propose that this work can be used as a first step towards building a null model for time series-based networks. In this respect, given that only about $2.5\%$ of the total occurrences present $|C|>0.875$, this threshold could be chosen to establish a link between two time series. Then, the statistics of the resulting network can be contrasted with those resulting from random walks to help discriminate a random origin of the networks. 

Finally, it would be interesting to asses the robustness of this kind of networks upon directed attacks. In particular, such a test in the negative slope regime of the degree distribution (quasi power-law, $H\approx 0.8$) could be compared against the results of the preferential attachment model that also yields a power law distribution.
\section{Acknowledgements}
Authors acknowledge financial support from CONACyT FRONTERAS 201. HP acknowledges postdoctoral fellowship from the same project and DGAPA-UNAM. We thank Francois Leyvraz for interesting discussions.
\section{Appendix A}
\label{Appendix A}
\begin{figure}
\includegraphics[height=2.1in,width=3.0in]{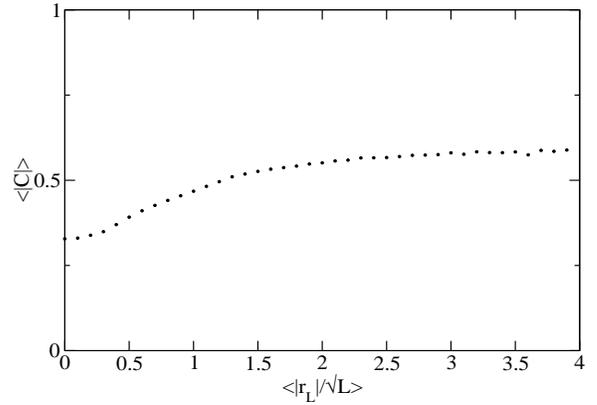}  
\caption{Average absolute correlation vs average $r_L/\sigma$ for $N=5000$, $L=500$, and number of samples=100.}\vspace{0.15cm}
\label{fig-av-abs-PCC-av-Rl1}
\end{figure}

\begin{figure}
\includegraphics[height=2.1in,width=3.0in]{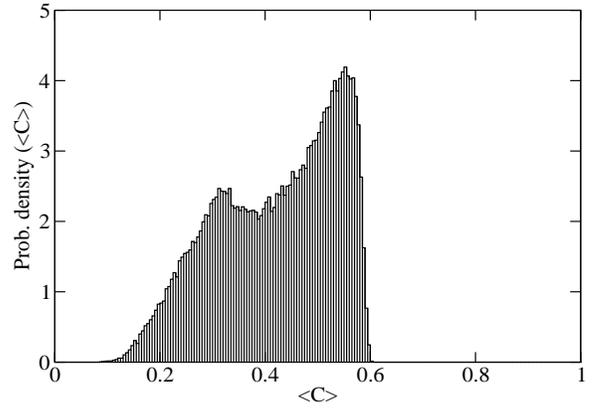}  
\caption{Average absolute correlation distribution for $N=5000$, $L=500$, and number of samples=100.}
\label{fig-av-abs-corr-distribution}
\end{figure}

Figure \ref{fig-av-abs-PCC-av-Rl1} shows that the average $<|C|>$ for walks with equal $|r_L|/\sigma$ varies more steeply for intermediate values of $|r_L|/\sigma$ ($1<|r_L|/\sigma<0.5$). This means that there are relatively greater overlaps in the $\rho(|C|)$s for nodes with $r_L/\sigma$ values on the two sides of the steepest region of the plot in Fig. \ref{fig-av-abs-PCC-av-Rl1}. The two peaks in the $<|C|>$ distribution for all the nodes, as shown in Fig. \ref{fig-av-abs-corr-distribution}, testifies the rough classification of these nodes into two groups. The larger overlap between $\rho(|C|)$ for different nodes leads to closer values of connection probability $p(H)$ for them. Hence, there exist two peaks in the degree distribution at intermediate values of $H$ corresponding to these two groups, and the weaker one disappears first as $H$ increases.  
\section{Appendix B}
\label{Appendix B}

In this appendix, we explain the shape of the degree correlation plots and the origin of discontinuities in the slopes of $<k_{nn}|k>$ and $<C(k)>$ vs $k$ plots.
When the basis of connectivity is PCC, which is a measure of similarity between the nodes, one would expect the neighbors of high degree nodes to have high degrees and \textit{vice-versa}, resulting in an overall assortative degree correlation as observed. However, this argument is not strictly valid in this case as the RWS used are not detrended. Due to the effect of trends, which is at the heart of the proportionality between $<C>$ and $|r_L|/\sigma$, PCC is not purely a measure of similarity between two RWS. In fact, there is a chance that two very similar RWS can have correlations equal to or higher than the correlation they have among themselves with a slightly dissimilar RWS, if that node has larger $r_L$. In other words, lack of similarity can be compensated or overcompensated by larger $r_L$ (the fitness) values to some extent. Clearly, if the two nodes having same or close $r_L$ value are already dissimilar, the effect of trends will dominate, and they both will be more likely to have higher correlations with a node with larger $r_L$ than them.  
Since the nodes for which $r_L$ is small are less likely to have similar RWS (due to a larger number of combinations of forward and backward steps as explained in section IIIC), they are more likely to be connected with nodes with larger $r_L$, hence with higher-degree nodes, than among themselves. Nodes with a larger value $r_L$ have more similarity in their RWS, and thus have a higher average correlation among themselves. Moreover,  $P(r_L)$ decays with $r_L$. As a result, the probability of having higher-degree neighbors decreases as $r_L$ increases, and hence with the degrees of the nodes, on average. This implies that $<k_{nn}|k>/k$ should decrease as $k$ increases, and accounts for the overall deviation from linearity observed. In addition to that, there is a discontinuity in the slope of the $<k_{nn}|k>$ vs $k$ plot near the value of $k$ corresponding to the left peak in the degree distribution (see for example Fig. \ref{fig-degree-dist}, H = 0.42). This basically means that the average behavior of all the nodes under a peak is similar to each other despite of the small differences in their degrees which results in a dip in the slope making it almost zero near the $k$ corresponding to the left peak.  This anomaly can also be associated with the discontinuity in the derivative of the $<CC>$ \textit{wrt} $k$, as it happens to be around the $k$ value corresponding to the left one of the two peaks in the degree distribution. Since the second peak in the degree distribution is closer to the maximum value of $k$, it does not lead to such a discontinuity.

\end{document}